# Designer Bloch Plasmon Polariton Dispersion in Hyperbolic Meta-Gratings

*Nicolò Maccaferri[1]*, Tommi Isoniemi[2,3], Michael Hinczewski[4], Marzia Iarossi[2,5], Giuseppe Strangi[4,6]*, and Francesco De Angelis[2]*

[1]Department of Physics and Materials Science, University of Luxembourg, L-1511 Luxembourg, Luxembourg

[2]Istituto Italiano di Tecnologia, I-16163, Genova, Italy

[3]Department of Physics and Astronomy, University of Sheffield, Sheffield S3 7RH, UK

[4]Case Western Reserve University, 44106 Cleveland, Ohio, USA

[5]Dipartimento di Informatica, Bioingegneria, Robotica e Ingegneria dei Sistemi (DIBRIS), Università degli Studi di Genova, I-16126 Genova, Italy

[6]CNR-Nanotec and Dipartimento di Fisica, Università della Calabria, I-87036 Rende, Italy

*nicolo.maccaferri@uni.lu; *gxs284@case.edu

**Abstract**

Hyperbolic metamaterials (HMMs) represent a novel class of fascinating anisotropic plasmonic materials, supporting highly confined propagating plasmon polaritons in addition to surface plasmon polaritons. However, it is very challenging to tailor and excite these modes at optical frequencies by prism coupling because of the intrinsic difficulties in engineering non-traditional optical properties with artificial nanostructures and the unavailability of high refractive index prisms for matching the momentum between the incident light and the guided modes. Here, we report the mechanism of excitation of high-k Bloch-like Plasmon Polariton (BPPs) modes with ultrasmall modal volume ($< \lambda/20$) using a meta-grating, which is a combined structure of a metallic diffraction grating and a type II HMM. We show how a 1D plasmonic grating without any mode in the infrared spectral range, if coupled to a HMM supporting high-k modes, can efficiently enable the excitation of these modes via coupling to far-field radiation. Our theoretical predictions are confirmed by reflection measurements as a function of angle of incidence and excitation wavelength. We introduce design principles to achieve a full control of high-k modes in meta-gratings, thus enabling a better understanding of light-matter interaction in this type of hybrid meta-structures. The proposed spectral response engineering is expected to find potential applications in bio-chemical sensors, integrated optics and optical sub-wavelength imaging.

## Introduction

Manipulation of photons at the nanoscale, well beyond the diffraction limit of light [1-4], has become a topic of great interest for the prospect of real-life applications [5], such as efficient energy harvesting and photosensitive chemical reactions [6-8], subwavelength waveguides [9], nanocavity lasers [10], opto-electronics [11], biochemistry [12] and nanomedicine [13]. Since the last decade conventional metallic materials have been molded with the most advanced nanofabrication techniques in order to create electromagnetically coupled nanostructured systems, dubbed metamaterials, with novel optical properties emerging from the subwavelength confinement of light [14]. Structured meta-surfaces enable an unprecedented control of the propagation direction of optical excitations on their surface. In this framework, hyperbolic metamaterials (HMMs) have received great attention from the scientific community due to their unusual properties at optical frequencies that are rarely or never observed in nature [15-17]. Furthermore, HMMs have been shown to enable negative refraction [18-21], resonant gain singularities [22], nanoscale light confinement [23], optical cloaking [24], as well as extreme biosensing [25,26], nonlinear optical phenomena [27], super resolution imaging and superlensing effects [28,29], plasmonic-based lasing [30], artificial optical magnetism [31], full control of decay channels on the nanoscale [32], etc. They display a hyperbolic iso-frequency surface [33-35], which originates from one of the principal components of their electric or magnetic effective tensor, having the opposite sign to the other two principal components. When considering the dielectric tensor, HMMs can be divided into two types: type I has one negative component in its permittivity tensor and two positive ones. In contrast, a type II HMM has two negative components and one positive. In practical terms type II appears as a metal in one plane and as a dielectric in the perpendicular axis, while type I is the opposite. Such anisotropic materials can sustain propagating modes with very large wave vectors and longer lifetime and propagation length in comparison to classic plasmonic materials [36] and exhibit diverging density of states [37], leading to a strong Purcell enhancement of spontaneous radiation [38,39]. Beyond the so-called natural hyperbolic materials, it is possible to mimic hyperbolic properties, for instance of type II, using a periodic stack of metallic and dielectric layers [40] that can support surface plasmons with large wave vectors [41] and whose effective permittivities for different polarizations have different signs [35,42].

## Results and discussion

In this work we focus on a specific type of artificial hyperbolic material, namely an HMM of type II made of alternating layers of Au and $Al_2O_3$, which has been coupled to a one-dimensional (1D) plasmonic diffraction grating, also known as plasmonic crystal (PC). A sketch of the meta-grating concept is presented in **Figure 1**. In the top panel, we have made a simple sketch of a 1D-PC made of gold, which is illuminated with a transverse magnetic (TM, p-polarized) light wave at an angle $\theta \neq 0°$. The experimental reflectance as a function of the wavelength of the incident light impinging at $\theta = 60°$ on a 1D-PC with grating period 450 nm and PMMA stripe dimensions of 250 nm (width) and 60 nm (thickness) with a 20 nm gold layer is plotted on the right. As can be seen by looking at the reflectance spectrum, no special features appear in the wavelength range 1000-1750 nm. Similarly, if we consider the experimental reflectance of an HMM crystal (middle-left panel) made of 8 alternating layers of Au (15 nm) and $Al_2O_3$ (30 nm) with on top an additional dielectric spacer (Al2O3, thickness 10 nm), we can see that no features are present in the same spectral range (middle-right panel). Furthermore, the sample is highly reflective in the NIR region, as already demonstrated in previous works on similar systems [43]. If we add the 1D-PC on top of our HMM crystal, we construct a meta-grating, and sharp and intense modes appear in the spectral region of interest. Moreover, the mode associated with the grating at 850 nm remains in the same position for the meta-grating.

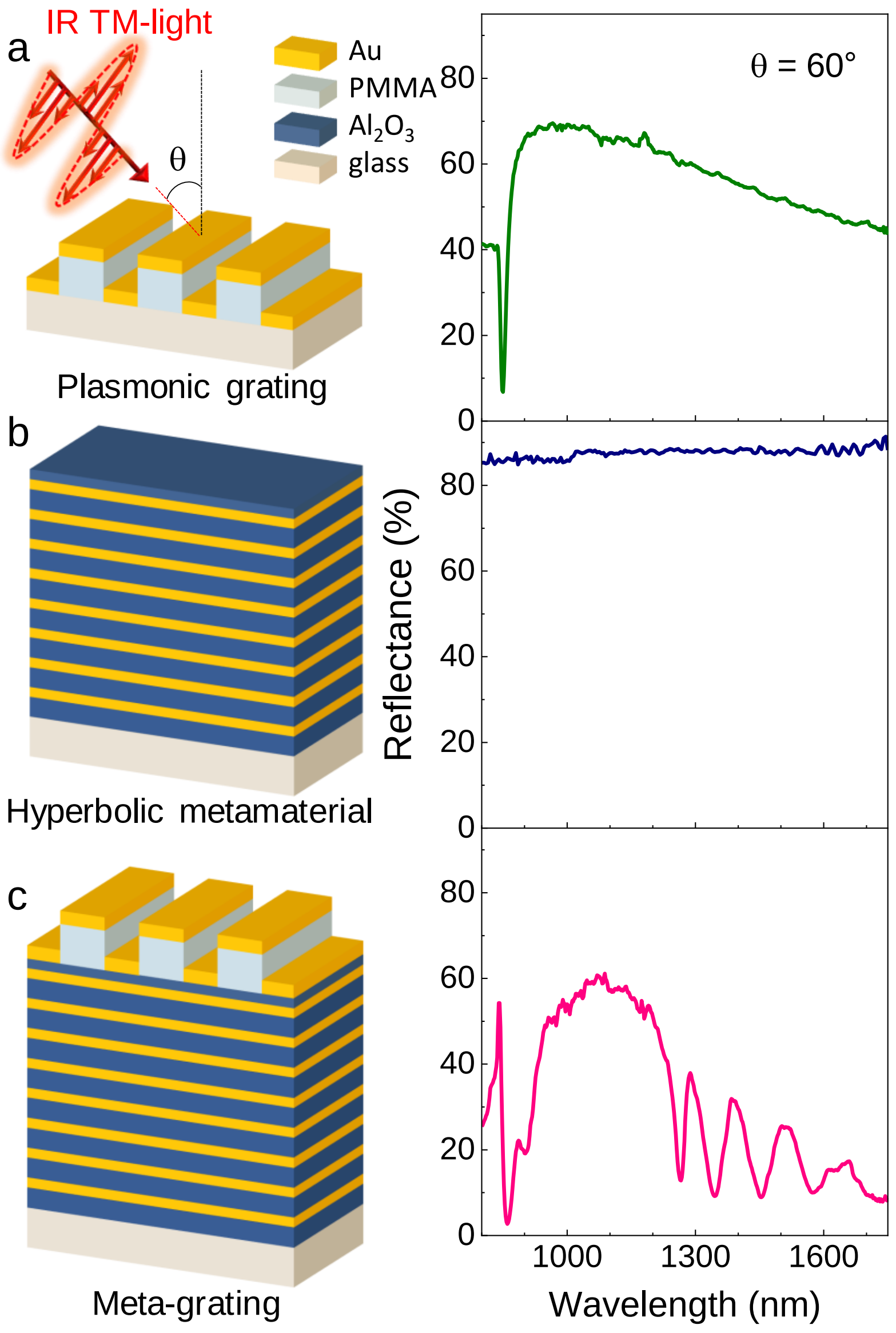

**Figure 1.** Sketches of (a) the control 1D-PC, (b) the HMM and (c) the meta-grating studied in this work. On the right side of each sketch, specular reflectance spectra measured with TM polarized light of the corresponding structures at incident angle θ = 60° are shown.

It is worth mentioning here that the 1D-PC does not support any diffraction-coupled plasmonic mode in the 1000-1750 nm range, as demonstrated by the experimental measurement reported in **Figure 1**. In straight contrast, the two systems, if coupled through a dielectric spacer, display an interesting behavior under the same excitation mechanism, namely a TM polarized light wave. It is important to note that with transverse electric (TE, s-polarized) incident light, no modes are observed in the meta-grating (see **Supporting Figure S1**). Although the 1D-PC itself does not support any propagating plasmon in this range of wavelengths, HMM can indeed support confined modes, which are not excitable by directly coupling the HMM with external radiation. To demonstrate this effect we have calculated, using a transfer matrix method (see Methods for additional details), the dispersion of the modes that can be excited in the HMM multilayer in the absence of a diffraction grating, **Figure 2a**. Under certain experimental conditions (such as the prism-coupling technique or high-energy electron beams) we can excite two Surface Plasmon Polariton (SPP) modes, indicated with S1 and S2, one at the interface with air and one at the interface with the glass substrate, respectively. Nevertheless, these modes fall outside the spectral region we are interested in. We observe other extremely confined and high-index propagating modes, which are within the multilayered structure known also as Bloch Plasmon Polaritons (BPPs), labelled BX, where X = 1, 2, 3, …, etc. These modes are the reported eigenmodes of an HMM multilayer [41]. We now consider the well-known dispersion relation of SPP in a grating, that is $k_{SPP} = \Lambda \sin\theta + mG$, where $\Lambda = 2\pi/\lambda$ ($\lambda$ is the wavelength of light), $G = 2\pi/a$ (a is the grating period), $\theta$ the angle of incidence and m is an integer number (positive or negative). If we calculate the geometrical dispersion of a grating with period 250 nm, and an angle of incidence of 60° for different values of m, for instance -1, 1 and 2, we end up with the red lines in **Figure 2a**. As it can be noticed, these curves intersect the BPP mode dispersion curves at precise energy values in the NIR spectral range, which we highlight with colored dots in **Figure 2a**. If we focus our attention on the grating mode (-1,0) we observe that it crosses the curves of the first four BPPs at 1370 nm (red dot – B1), 1520 nm (green dot – B2), 1680 nm (orange dot – B3) and 1840 nm (purple dot – B4). These results have also been confirmed through a scattering matrix method [44]. By considering also the grating structure on top of the HMM multilayer, it results in the blue curve in **Figure 2b**, reproducing significantly well the experimental curve reported in **Figure 1c**. The slight difference between the position of the dips, as well as the relative spectral separation between them, is due to the difference in the values of the dielectric constants of the deposited materials and those used for the calculations [45,46]. Nevertheless, it is clear that our theoretical results and our argumentations (grating dispersion coupled to BPP dispersion) confirm the idea that the presence of a plasmonic diffraction grating on top of an HMM multilayer is responsible for the excitations of these modes in the NIR. We also analyzed in detail the various ways the different diffraction orders can excite BPP modes. For instance, the B1 mode at 1370 nm and the B3 mode at 1680 nm are examples of the simplest scenario: one diffraction order is dominant throughout the interior of the HMM (see the (-1,0) curve in **Figures 2d** and **2f**). The mode identity is given by where this diffraction order intersects the modal lines in **Figure 2a**. On the contrary, the mode we call B3 at 870 nm is a bit more complicated, because while (-2,0) is the largest contribution throughout most of the HMM, there is also a significant (1,0) contribution. Thus, we get a mode with roughly the characteristics of B3 (compare **Figure 2e** and **2g**), based on the shape of the Poynting vector distribution (see the negative $S_x$ peaks in the alumina layers in **Figure 2g**), but mixed with counter-propagating modes (positive peaks) excited via (1,0).

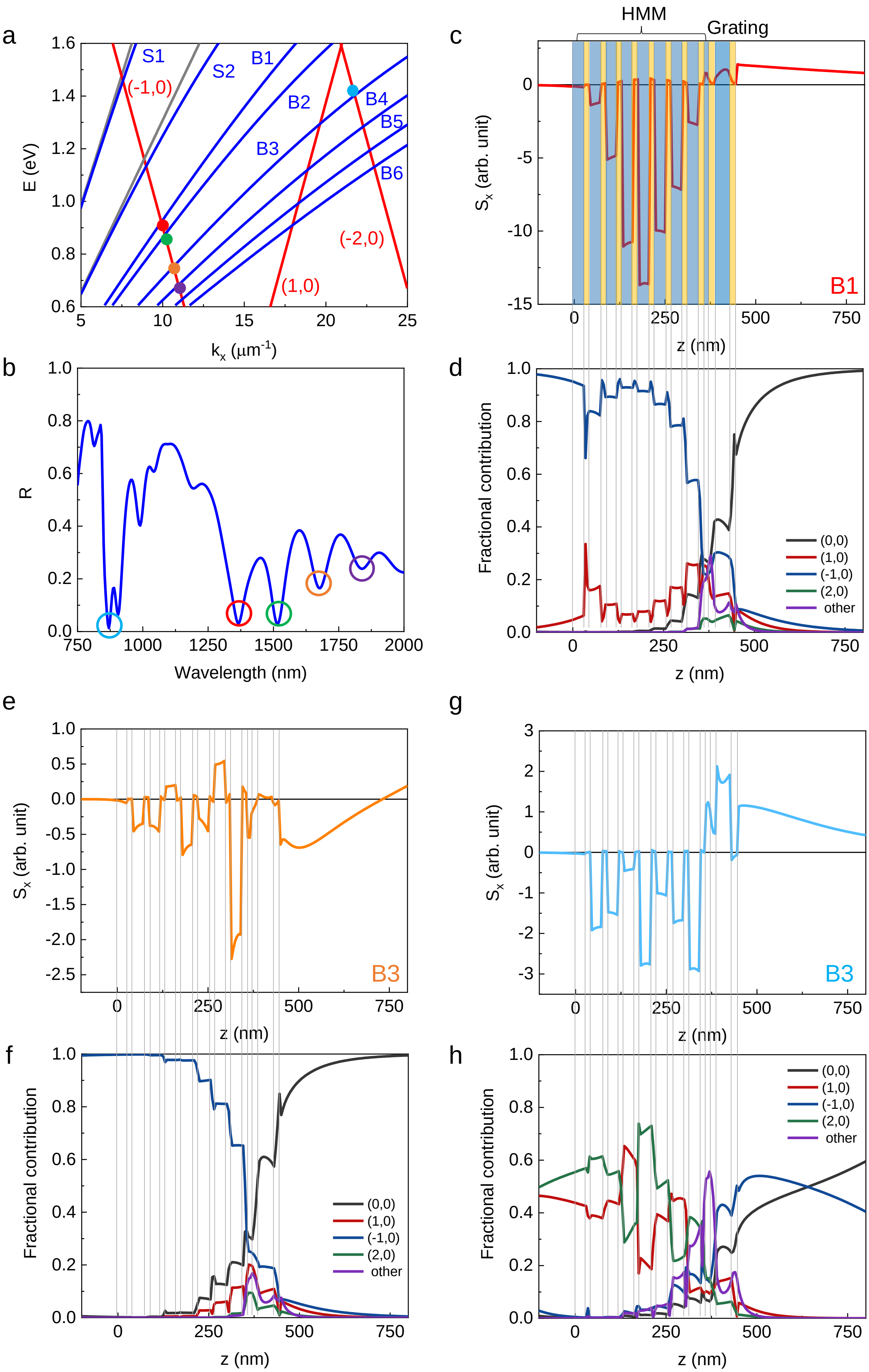

**Figure 2.** Results of Fourier modal calculations for the meta-grating system. a) Modal dispersion diagram in terms of incident light energy E versus transverse wavevector $k_x$. Thin blue lines are modes calculated for the HMM structure without the grating using the optical transfer matrix approach. The modes are labeled as follows: Sn (n=1-2) are SPPs, Bn (n=1-6) are BPPs. The dashed red lines correspond to energy-wavevector relations for different diffraction orders: (-1,0), (1,0) and (-2,0). The intersection of these relations and the energy at each minimum highlighted in panel b is shown as a colored dot. Note the dots do not perfectly overlap with the calculated mode lines, because the actual mode positions are shifted slightly by the presence of the grating. Curves corresponding to propagation in air and glass are drawn in gray. b) Reflectance versus wavelength. Three BPP modes (B1 at 1370 nm, B2 at 1520 nm, B3 at 870 and 1680 nm, and B4 at 1840 nm) are highlighted. c), e), g) For each of the modes highlighted in panel b, the transverse Poynting vector component versus height within the structure and d), f), h) the corresponding fractional contribution of each diffraction order to the field within the structure. The dominant contributions are from orders (0,0), (1,0), (-1,0), and (-2,0). All others are summed together and shown as "other".

In this geometry and spectral range, B3 was the clearest example of the same BPP mode that can be excited via two different diffraction orders at two different wavelengths (see **Figures 2a, 2e-h**). Finally, it is worth noticing that there is also an additionally (-1,0) contribution from the grating mode close to the multilayer/air interface, which might come from the SPP mode excited by the grating and which is clearly present in the experimental curve shown in **Figure 1a**. This is the mode at around 900 nm close to the B3 mode at 870 nm.

A more detailed mode analysis unveils that actually the geometrical diffraction mode (-1,0) supported by the 1D-PC, upon excitation with the far-field TM radiation, couples with the BPP modes supported by the HMM at certain wavelengths. In **Figure 3a** we show a representative SEM image of a FIB-milled cross section on the experimentally realized 1D-PC introduced on top of a HMM.

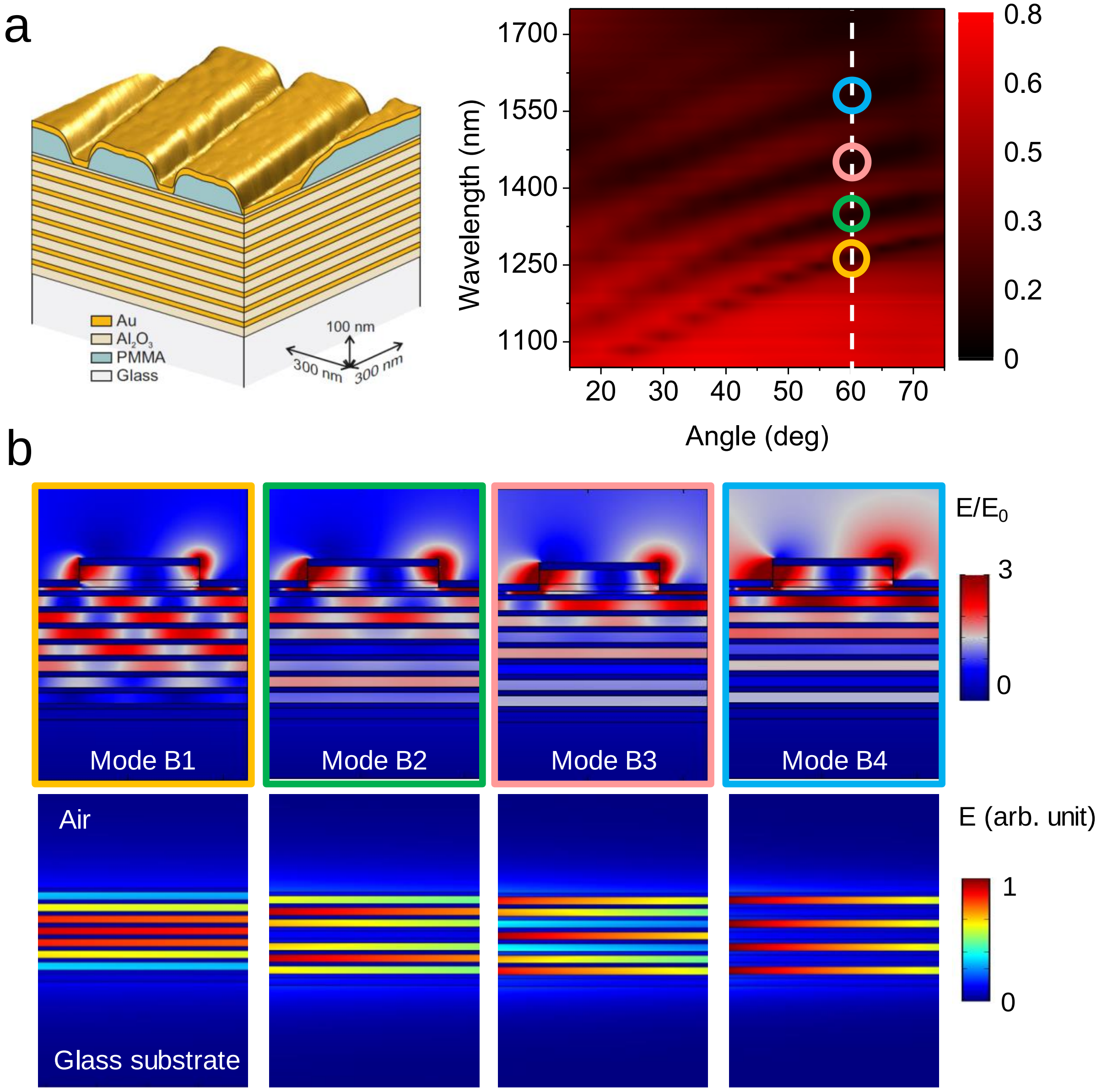

**Figure 3.** (a) Colorized scanning electron micrograph of a FIB trench on the 1D-PC made of PMMA (60 nm thick) on a HMM multilayer covered with 20nm of Au. (b) Experimental reflectance as function of the wavelength and the incident angle of the incoming TM-polarized light. The four BPP modes excited in the meta-grating at $\theta = 60°$ are highlighted with colored circles. (c) Near-field distribution of the four BPP modes which are excited in the meta-grating (top panel) and exact solution of the electromagnetic field distribution of a HMM on a glass substrate without the 1D-PC on top obtained at the same energies with a finite element method-based mode analysis (bottom panel).

The reflectance spectrum of the meta-grating as a function of the wavelength and the incident angle of the impinging TM-polarized far-field radiation is plotted in **Figure 3b**. Four dips are clearly visible in the studied spectral range. These modes are also highlighted with colored circles at $\theta = 60°$ (for more details about the fabrication of the samples and the optical experiments see Methods). In the top-panel of **Figure 3c** we plot the near-field distribution of the four modes excited in this spectral range $\theta = 60°$. As can be inferred by looking at the field distributions within the HMM plotted in the top panel **Figure 3c**, particular field distributions appear at the wavelengths where the grating mode crosses the BPP modes. These are exactly the field distributions we can expect by solving numerically the eigenvalues problem using a finite element method-based mode analysis (see also Methods) of a multilayered HMM on a glass substrate without the 1D-PC on top of it (bottom-panel of **Figure 3c**).

This proves that, although no SPP modes are supported by the 1D-PC in this wavelength range, the fact that a diffraction order is present allows the excitation of BPP modes which are actually impossible to excite with direct coupling to far-field infrared radiation. It might be that localized plasmons are excited within the 1D-PC through a diffraction mechanism. If we look at **Figure 3c**, we can also observe that the field distribution at the grating level is almost the same for all four the cases, indicating that i) it is only one grating mode which is actually contributing to the excitation of the BPP modes and ii) the mode is almost localized.

To prove the latter hypothesis, we have also calculated the reflectance of systems where we considered different type of gratings on top of the HMM. In **Figure 4a** we plot the four geometries considered, along with the related calculated reflectance spectra at $\theta = 60°$ and the near-field distributions for the B2 mode.

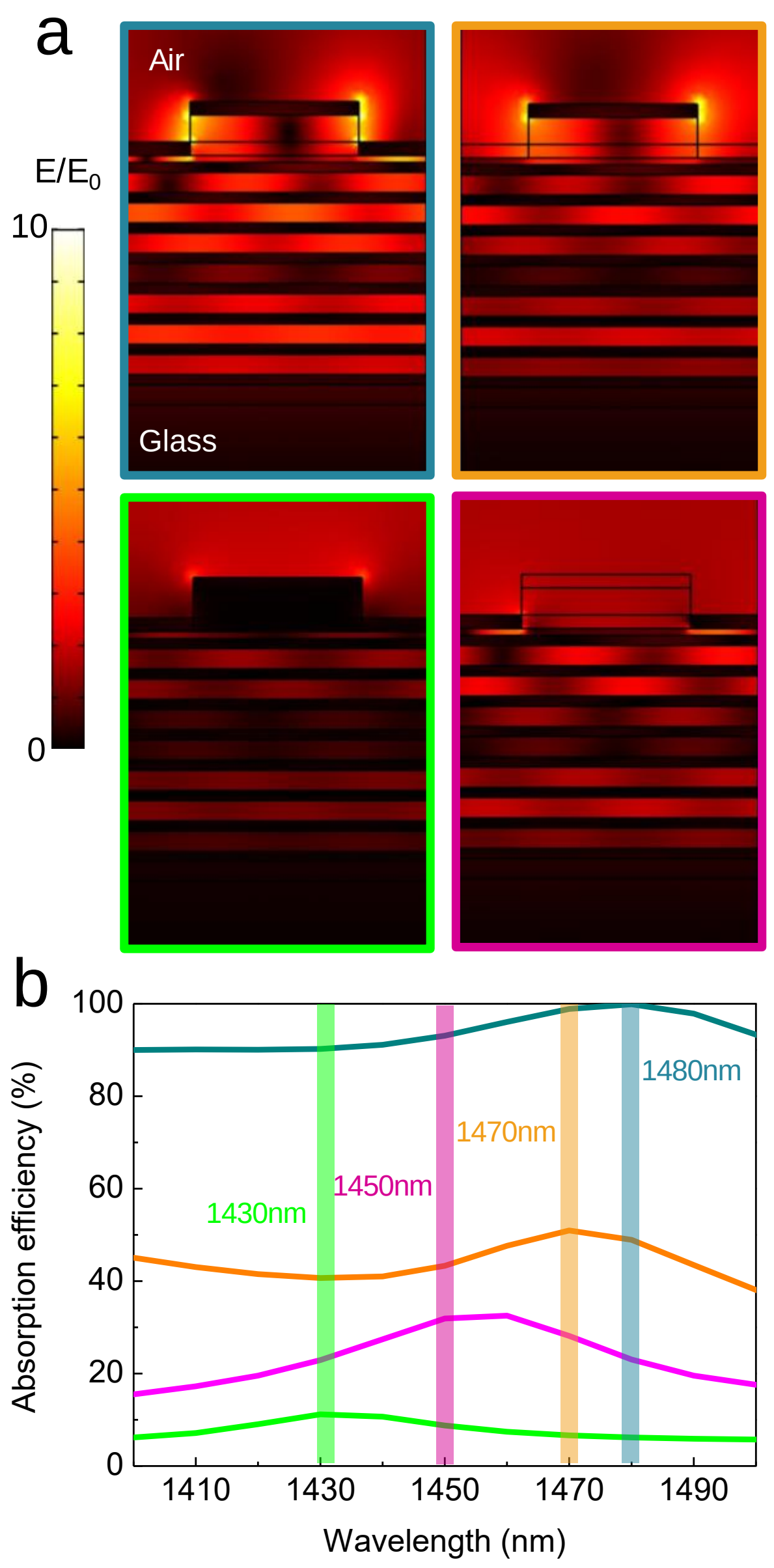

**Figure 4.** (a) Calculated near-field distribution of the B2 mode excited in four different meta-gratings: experimental case (top-left), namely a PMMA stripe with of top a gold thin film (thickness 20 nm), a gold stripe on top of a PMMA bar, and no gold on top of the dielectric spacer (top-right), a pure gold grating (bottom-left) and an anti-stripe (bottom-right), namely a gold film with a stripe etched on it. The excitation wavelengths of the B2 mode in each case are marked in (b). (b) Corresponding calculated absorption efficiency for the four cases reported in (a).

By directly comparing the response of the four meta-grating cases presented in the **Figure 4**, it turns out that it is actually a localized mode at the thin Au stripe edges the main responsible for the excitation of the BPP modes. In particular by comparing the responses of the Au grating on top of PMMA stripes with a dielectric spacer (simulation of the experimental case – top-left panel) and the case of the Au layer on top of the PMMA stripe (top-right panel). If we consider an ideal grating, namely a grating made of pure gold (bottom-left panel), or an Au thin film with a stripe etched within it (anti-stripe case, bottom-right panel) we can certainly excite the B2 mode but with much less efficiency. This is also quite clear by looking at **Figure 4b**, where we plot the related calculated absorption efficiency spectra in the wavelength range where the B2 mode is excited using these four different diffraction gratings. By playing with the grating composition, one can pass from a rather negligible absorption (pure gold grating case) to a huge absorption (> 99%) in the case of the Au film deposited on top of a PMMA stripe, which is the same of our experimental case, although there we reach an absorption efficiency of ≈ 90% due to structural defects. Similar effects were also recently studied in the visible spectral region by Azzam et al., who observed the formation of very sharp and high-quality factor hybrid photonic-plasmonic modes when coupling a 1D-PC with a dielectric optical waveguide [47]. In that study the plasmonic propagating modes can be excited directly by coupling with far-field radiation and hybridize with the cavity modes in the waveguide, while in our case it is crucial to couple the two systems, which are not displaying any bright mode if taken separately.

We have also studied how the thickness of different building blocks (either the HMM multilayer or the PMMA/Au grating on top), such as the PMMA bar, the $Al_2O_3$ spacer or the metal/dielectric layers, can affect the dispersion of the modes.

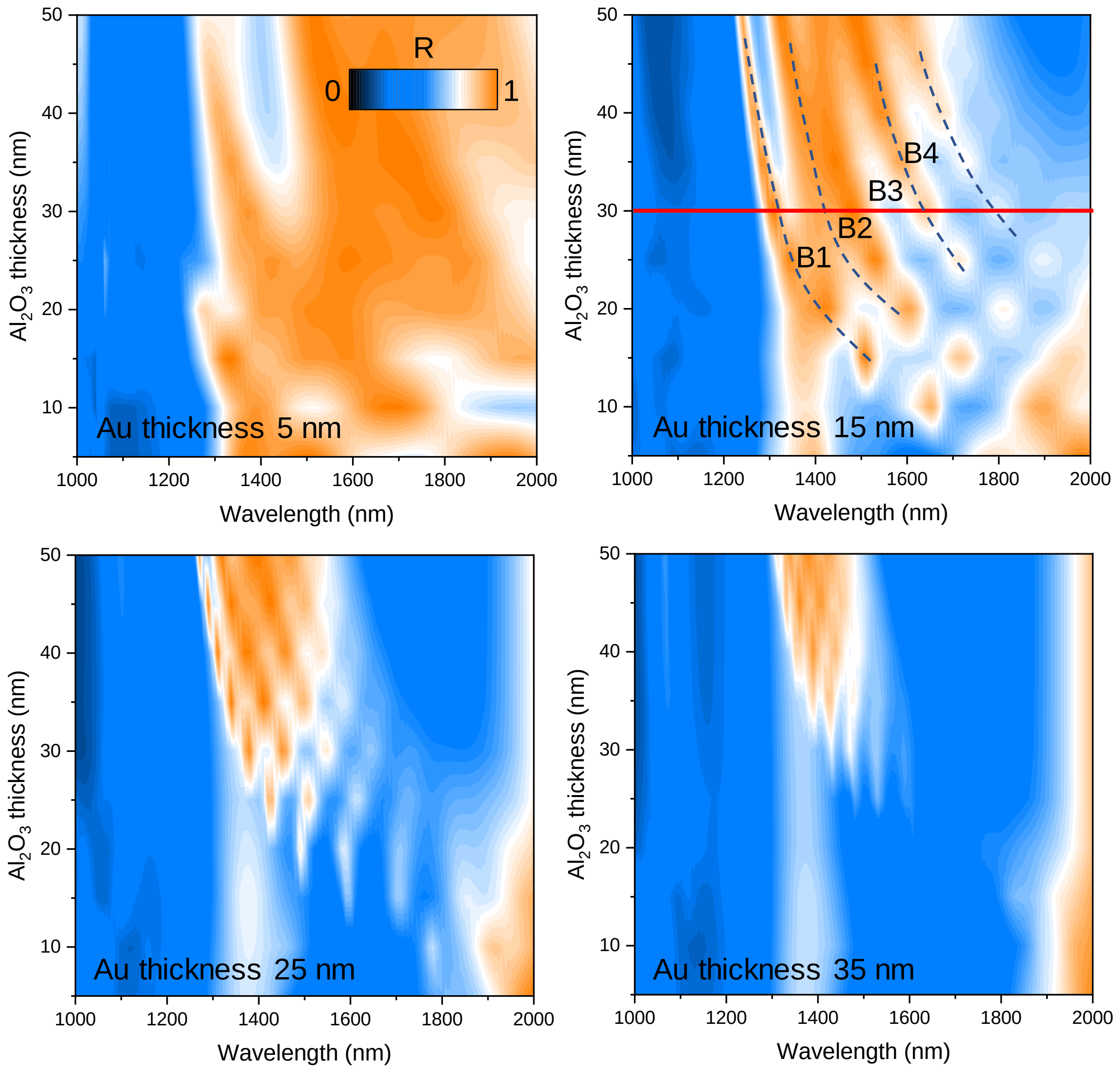

**Figure 5.** Calculated reflectance of an meta-grating as a function of the $Al_2O_3$ thickness in the repeated multilayer and the wavelength of the impinging light, and where the thickness of the Au layers has been changed from 5 nm to 35 nm. The red continuous line indicate the experimental case, and the black dashed lines are guide for the eyes to help the reader visualizing the BPP modes reflectance dependence on the wavelength and $Al_2O_3$ thickness.

In **Figure 5** we display the reflectance of the meta-grating studied experimentally as a function of the wavelength of the incident light (for $\theta = 50°$) and of thickness of the $Al_2O_3$ layers for four different values of the thickness of the Au layers. As it can be seen there is a threshold dictated by the experimental value of 15 nm, above which four BPP modes merge and disappear. Moreover, already for an Au thickness of 25 nm, the four Bloch modes strongly red-shift and disappear for values of the $Al_2O_3$ thickness below 30 nm.

We then calculated the dependence of the BPP modes dispersion on the thickness of the PMMA bar. We left the Au film thickness constant and equal to 20 nm in all the cases. In **Figure 6** we report the reflectance of the meta-grating for θ = 50° and for different values of the PMMA bar thickness, as a function of the wavelength of the incident light and of the $Al_2O_3$ spacer thickness. As it can be inferred, to achieve a proper coupling between the metallic grating and the HMM multilayer, the PMMA bar thickness has to be larger than the Au film thickness, that is, we need to separate well the top part of the grating from the bottom part. Indeed, only for thicknesses above 40 nm we start to see the appearance of the four BPP modes reported in **Figure 1**. It is also interesting to note that for PMMA bar thicknesses above 60 nm (as in the experimental case), the dispersion of the four BPP modes is not strongly modified. Moreover, it is also important to note that the $Al_2O_3$ spacer thickness should not be larger than 20-30 nm, because for thicknesses beyond this threshold the reflectance dips become less intense, which is a signature that the coupling between the grating and the HMM multilayer is not optimal.

It is also important to note that the BPP modes supported by the HMM display different refractive indices depending on the order of the mode [41]. These indices are known as modal indices, and affect both the propagation and penetration distance of a surface wave, in our case the BPP modes. In **Figure 7a** and **7b** are reported the real and imaginary part, respectively, of the modal index of the first seven modes supported by the HMM. For clarity we plot also the same quantities for the two SPP modes at the air and glass interfaces (blue and yellow lines, respectively).

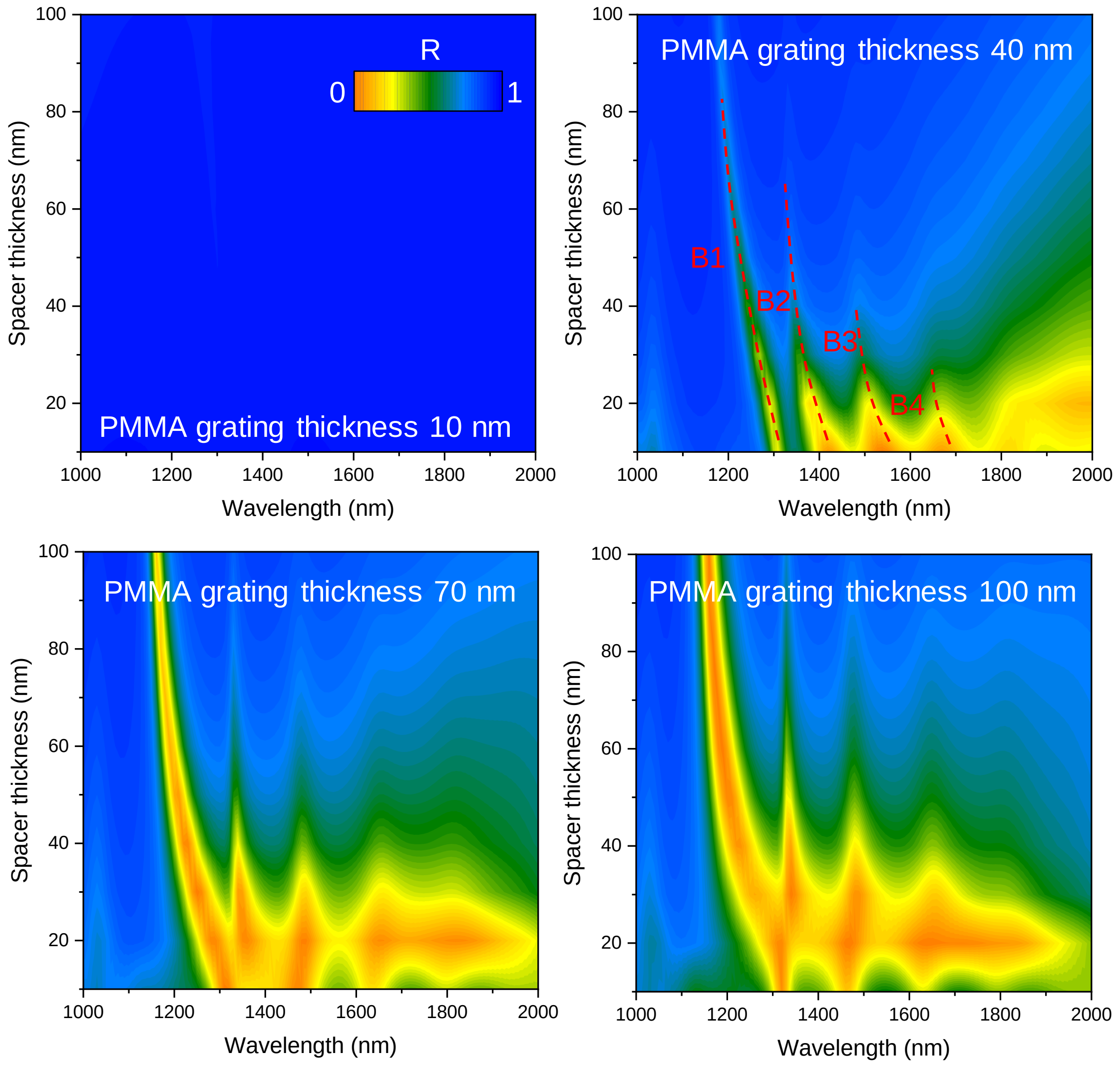

**Figure 6.** Calculated reflectance of an meta-grating as a function of the $Al_2O_3$ spacer thickness and the wavelength of the impinging light, and where the thickness of the PMMA bar has been changed from 10 nm to 100 nm. The red dashed lines are guide for the eyes to help the reader visualizing the BPP modes reflectance dependence on the wavelength and $Al_2O_3$ spacer thickness.

As the number of the mode increases, also their losses (plotted in **Figure 7c**) increase, thus allowing to slow down radiation velocity (see **Figure 7d**, where the group velocity of the modes is plotted) within the structure. This is important in view of practical photonic applications where tunable life-times (thus losses) can be exploited in the system, and our architecture might represent an interesting candidate to achieve this functionality. Finally, it is worth mentioning here that high-index and slow photonics modes like these can be used for enhancing nonlinear interactions as well as in optical circuitry, especially for buffering, switching and time-domain processing of optical signals [48].

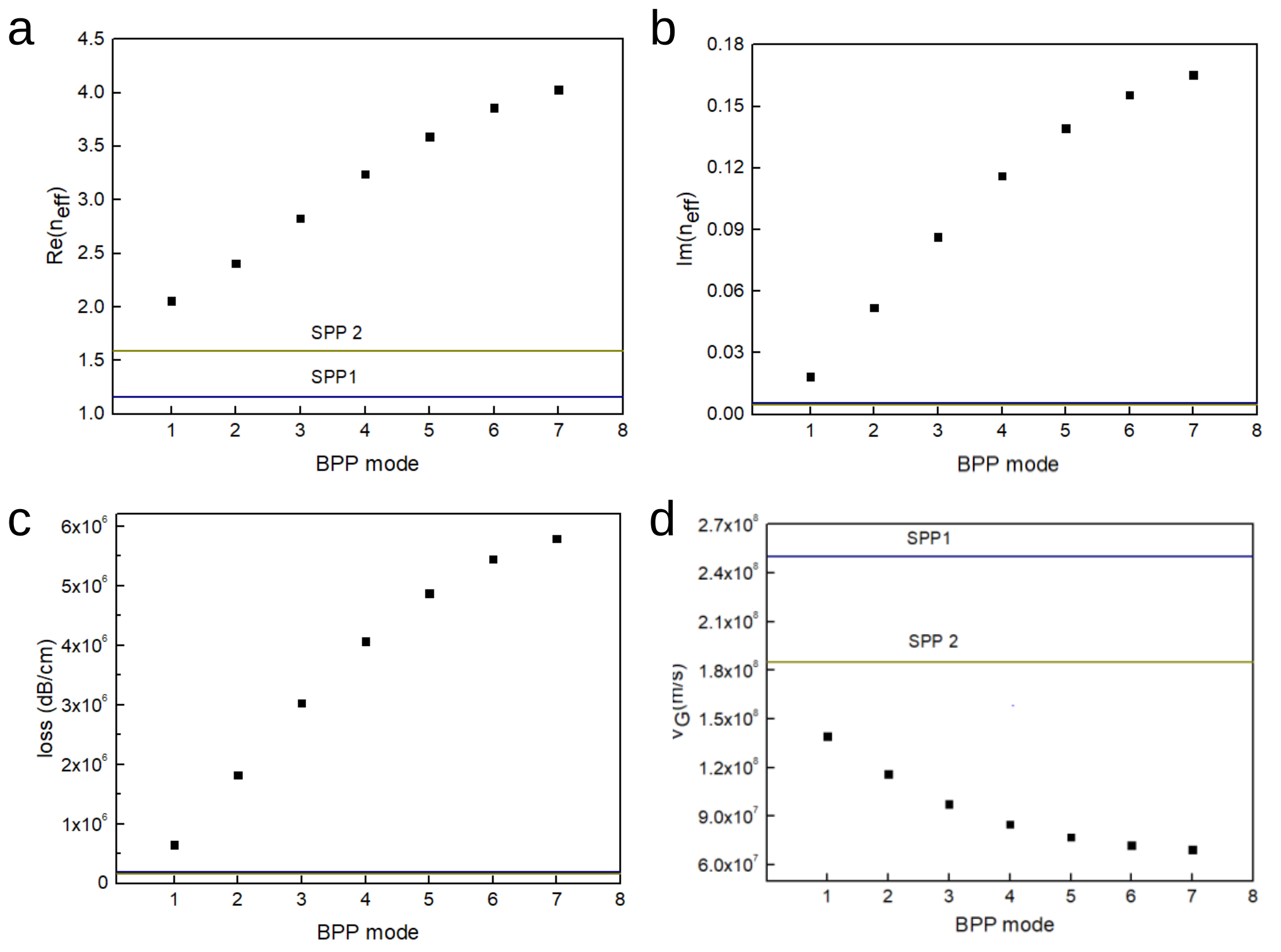

**Figure 7.** The real (a) and imaginary (b) parts of modal indices for the BPP modes in a 8-bilayer HMM. (c) Losses and (d) group velocity for the BPP modes.

## Conclusion

We have demonstrated the excitation mechanism of Bloch Plasmon Polariton modes in the radiation continuum using a meta-grating approach. Due to the strong coupling of the hyperbolic and the plasmonic modes, the nature of the modes is exchanged at given spectral points, giving rise to sharp modes with geometry-tailored quality factor and tunable spectral position. These modes are characterized by high absorption efficiency (>99%) in the near-infrared spectral region. The system studied exhibits an exquisite set of phenomena including the formation of far-field radiation coupled high-index propagating modes with tunable life-times and group velocities, photonic bandgap engineering and slow light with broad spectral robustness. We believe that these hybrid structures are perfect candidates for applications in tunable-threshold lasers, sharp spectral filters, perfect absorbers, enhancement of nonlinear phenomena as well as biochemical sensors.

METHODS

*Sample fabrication:* The hyperbolic metamaterial samples were produced on a 1 mm thick glass slide (float glass, BK7 equivalent) cleaned with acetone, isopropanol and oxygen plasma. A 30 nm thick layer of aluminum oxide, confirmed with ellipsometry, was produced with an atomic layer deposition (ALD) process using trimethylaluminum and oxygen plasma as precursors at 80 °C in an Oxford Instruments FlexAL reactor. The sample was transferred with a vacuum shuttle to a electron beam evaporation chamber (Kenosistec KE300ET), and a gold layer with a thickness of 15 nm as measured by ellipsometry and atomic force microscopy was deposited at room temperature. These two deposition processes were repeated eight times to produce an alternating multilayer on the glass, with an additional 10 nm ALD-produced $Al_2O_3$ as the final layer for the HMM. By using the shuttle, the sample was kept continuously in vacuum between the deposition processes to prevent the introduction of impurities between the layers. The grating on the multilayer was produced by electron beam lithography (EBL), starting with a spin coated 60 nm polymethylmetacrylate (PMMA 950) resist layer, which was exposed using a Raith 150 Two EBL system with a 20 kV beam to expose grating lines with a 450 nm period. The resist was developed using a mix of 1 part methylisobutylketone and 3 parts isopropanol for 40 s. Subsequently 20 nm of gold was evaporated (KE300ET) on the sample with neodymium permanent magnets near the sample to prevent charged particles from reaching the surface. No lift-off was done, leaving the PMMA layer as part of the grating. The corresponding control sample was made on a similar glass slide with a 20 nm indium tin oxide (ITO) layer produced with a Kenosistec KS500C sputter coater at room temperature using an ITO target at 250 °C with 1 sccm $O_2$ flow in the chamber. Subsequently PMMA coating, EBL and 20 nm Au coating were performed identically as with the hypergrating sample. The samples were also characterized by scanning electron microscopy after FIB milling with a 30 kV gallium ion beam using a FEI Helios NanoLab 650 dual beam system.

*Optical measurements:* Spectroscopic ellipsometry with a J.A. Woollam V-VASE system was done to acquire measurements for fitting refractive index functions and thicknesses of single layers to be used for simulations. Reflection spectra of the meta-gratings and control samples were taken with the same system. Specular reflectance spectra had typically a wavelength range of 270 nm to 1800 nm with a resolution of 2 nm, and were measured sequentially with a xenon lamp and a monochromator. The spectra were acquired with a focused probe accessory with lenses for both incident and reflected beams from areas less than 0.5 mm in diameter.

*Simulation details:* A homemade transfer matrix method code has been used to calculate the mode dispersion in **Figure 2a**, while the full optical response has been calculated by using the scattering matrix method [44] by considering the full structure and also by using the finite element method implemented in Comsol Multiphysics (RF module, 2D model geometry).

ACKNOWLEDGMENTS

NM acknowledge support from the FEDER Program (grant n. 2017-03-022-19 Lux-Ultra-Fast) and by the Luxembourg National Research Fund (CORE Grant No. C19/MS/13624497 'ULTRON').

## SUPPORTING INFORMATION

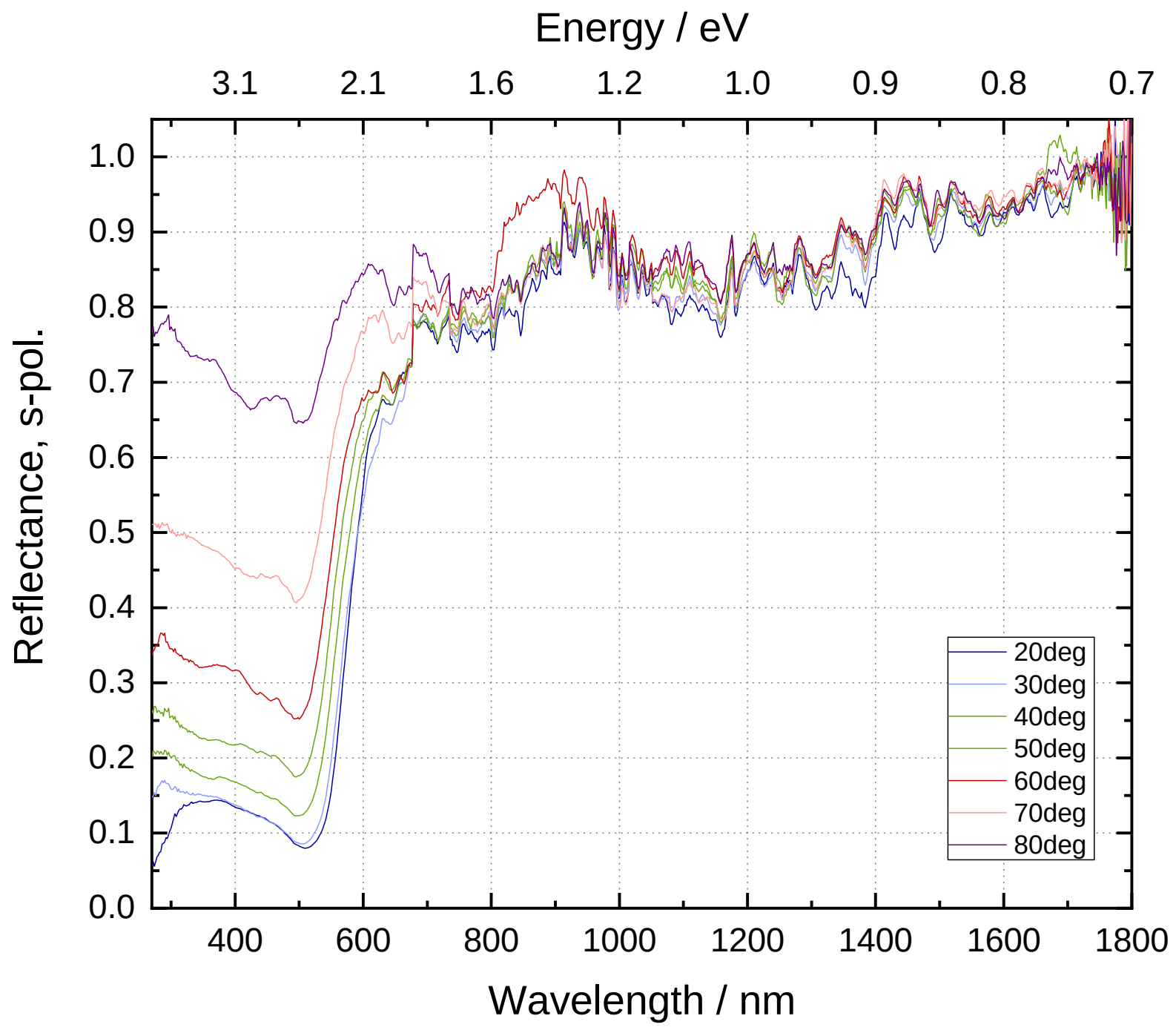

**Figure S1.** Specular reflection spectra at TE polarization measured from a meta-grating with 250 nm wide PMMA stripes. Incident angles from 20° to 80° in relation to sample normal.